\documentclass{article}

\usepackage{amssymb,amsfonts,amsthm}
\usepackage[cp1251]{inputenc}
\title{Chosen-ciphertext attack on noncommutative Polly Cracker}
\author{S.Bulygin}

\begin{document}
\maketitle
    \section{Noncommutative Polly Cracker and preliminaries from noncommutative algebra}
    The noncommutative Polly Cracker cryptosystems were developed by T.Rai in his Ph.D. dissertation ([1]), and rely
    on the fact that there are ideals of noncommutative algebras over finite fields that have infinite reduced Groebner
    bases.
    \\
    \indent
    First let us briefly present notations that will be used further in the text. Everything in this section
    is based on [1].We will be working with a
    noncommutative algebra $\mathbb{F}_q<X>$, where $X=\{x_1,\dots,x_n\}$, which is an algebra of noncommutative
    polynomials. By a monomial, we mean a finite noncommutative word in the alphabet $X$.
    We use the letter $B$ to denote the set of monomials.
    We define multiplication in the set $B$ of monomials by concatenation.
    The next important thing is the notion of an admissible ordering.
    A well-ordering $>$ on $B$ is said to be \emph{admissible} if it satisfies the following conditions for
    all $p, q, r, s \in B$:
    \begin{itemize}
            \item if $p < q$ then $pr < qr$;
            \item if $p < q$ then $sp < sq$ ;
            \item if $p = qr$ then $p > q$ and $p > r$.
      \end{itemize}
    \indent
    Let $>$ be an admissible ordering on the monomials and $f\in\mathbb{F}_q<X>$. We say
    that a monomial $b_i$ occurs in $f$ if the coefficient of $b_i$ in $f=\sum \alpha_i b_i$ is not zero.
    We say that $b_i$ is the \emph{tip} of $f$, denoted $tip(f)$, if $b_i$ occurs in $f$ and $b_i\ge b_j$ for all $b_j$
    occurring in $f$. We denote the coefficient of $tip(f)$ by $Ctip(f)$.
    If $S\subseteq\mathbb{F}_q<X>$, then we write $Tip(S) = \{b \in B : b = tip(f)$ for some nonzero $f \in S\}$ and
    $NonTip(S) = B - Tip(S)$.
    \\
    \indent
    Another thing we need is the notion of division of a polynomial $g\in\mathbb{F}_q<X>$ by polynomials $f_1,\dots,
    f_k\in\mathbb{F}_q<X>$. To perform such a division means to find nonnegative integers $t_1, t_2, \dots , t_k$ and
    elements $u_{ij}, v_{ij}, r \in \mathbb{F}_q<X>$, for $1 \le i \le k$ and $1 \le j \le t_i$ such that:
    \begin{itemize}
        \item $g =\sum_{i=1}^k\sum_{j=1}^{t_i} u_{ij} f_i v_{ij} + r$;
        \item $tip(g) \ge tip(u_{ij}f_i v_{ij})$ for all $i$ and $j$;
        \item $tip(f_i)$ does not divide any monomial that occurs in $r$, for $1 \le i \le k$.
    \end{itemize}
    Note that if $r \ne 0$, then $tip(r) \le tip(g)$; r is the \emph{remainder} of the division.
    \\
    \indent
    On notions of a Groebner basis in noncommutative case cf. [1].
    \\
    \indent
    Now we present the noncommutative Polly Cracker from [1]. It can be summarized as follows.
    \\
    \indent
    \emph{Private Key:} A Groebner basis, $G =\{g_1, g_2, \dots, g_t\}$ for a two-sided ideal, $I$, of a noncommutative
    algebra $\mathbb{F}_q<X>$ over a finite field of $q$ elements.
    \\
    \indent
    \emph{Public Key:} A set, $B = \{ q_r : q_r =\sum_{i=1}^t\sum_{j=1}^{d_{ir}}f_{rij}g_i h_{rij}\}_{r=1}^s
    \subseteq I$, chosen so that computing a Groebner basis of $<B>$ is infeasible.
    \\
    \indent
    \emph{Message Space:} $M = NonTip(I)$ or a subset of $NonTip(I)$.
    \\
    \indent
    \emph{Encryption:} $c = p + m$, where $m \in M$ is a message and $p =\sum_{i=1}^s\sum_{j=1}^{k_i}
    F_{ij}q_i H_{ij}$ is a polynomial in $J = <B>\subseteq I$.
    \\
    \indent
    \emph{Decryption:} Reduction of $c$ modulo $G$ yields the message, $m$.
    \\
    \indent
    Note that for practical reasons T.Rai proposes to use   $G$ containing only one element $g$.

    \section{Cryptanalysis of noncommutative Polly Cracker}
    In [3] and [4] it was shown that (commutative) Polly Cracker (first proposed in [2]) and its various modifications
    are susceptible to chosen
    ciphertext attacks. We will now show that noncommutative Polly Cracker is also susceptible to a chosen-ciphertext
    attack.
    In fact, we will
    only need one "fake" ciphertext in order to be able to decrypt all further ciphertexts correctly.
    In the following we assume that we know the form of $g$ (e.g. $g=\alpha xy+\beta x+\gamma y+\delta$,
    where $\alpha, \beta, \gamma, \delta \in \mathbb{F}_q$, cf. for example section 5.1.3 of [1]).
    \\
    \indent
    The main idea relies on the following observation. Let $I=<g>$, and consider $tip(g)$. We have:
    \begin{displaymath}
        tip(g)=Ctip(g)^{-1}g-Ctip(g)^{-1}\cdot tail(g),
    \end{displaymath}
    where $tail(g)=g-Ctip(g)\cdot tip(g)$. Note, that $tip(g)$ does not divide any monomial in $tail(g)$. This means
    that $-Ctip(g)^{-1}\cdot tail(g)$ is the remainder of division of $tip(g)$ by $g$, or equivalently, it is the result
    of decryption of the "fake" ciphertext $tip(g)$.
    \\
    \indent
    Now, let us go on to the chosen-ciphertext attack itself.
    Let us construct a "fake" ciphertext $c'=t\cdot tip(g)\cdot s + \sum F_{ij}q_i H_{ij}$,
    where $t,s\in\mathbb{F}_q<X>$ are such that any monomial of $t\cdot tail(g)\cdot s$ is not divisible by $tip(g)$.
    Polynomials $t$ and $s$ are chosen for masking the "fake" ciphertext and, in principle, can be dropped out.
    We have:
    \begin{displaymath}
        t\cdot tip(g)\cdot s=Ctip(g)^{-1}t\cdot g\cdot s-Ctip(g)^{-1} t\cdot tail(g)\cdot s,
    \end{displaymath}
    and using this latter assumption we obtain that $-Ctip(g)^{-1}t\cdot tail(g)\cdot s$ is the remainder
    of division of $t\cdot tip(g)\cdot s$ by $g$, and thus this is the remainder of division of $c'$ by $g$
    (as $\sum F_{ij}q_i H_{ij}$ reduces to 0 modulo $G=\{g\}$).
    \\
    \indent
    A next simple example shows that requirements on $t$ and $s$ can be easily satisfied. For instance, let us take
    $g=x_1\cdot\dots\cdot x_6+c_1 x_1+\dots +c_6 x_6+c_0, c_0,\dots,c_6\in\mathbb{F}_q\setminus\{0\}$ as in section
    5.1.2 of [1]. Then $tail(g)=c_1 x_1+\dots +c_6 x_6+c_0$ (under any admissible ordering) and we can take
    $t:=x_2 x_4+x_2 x_3 x_6+x_4 x_1 x_5; s:= x_5 x_1 x_3+ x_6 x_2 x_4$. One easily sees that no monomial of
    $t\cdot tail(g)\cdot s$ is divisible by $tip(g)=x_1\cdot\dots\cdot x_6$. It is also clear that many more
    variants of $t$ and $s$ can be proposed.
    \\
    \indent
    So, going back to our construction we see that if we send a "ciphertext" $c'$, we obtain a "plaintext"
    $p'=-Ctip(g)^{-1} t\cdot tail(g)\cdot s$. We know $t$ and $s$, so we can easily deduce
    $-Ctip(g)^{-1}\cdot tail(g)$ from $p'$. Now construct $g'=tip(g)+Ctip(g)^{-1}\cdot tail(g)$. We have
    $Ctip(g)\cdot g'=g$, so $I=<g>=<g'>$, and thus we can use $g'$ in order to decrypt ciphertexts
    to correct plaintexts, which is equivalent to knowing the private key $G=\{g\}$. Indeed, if for a ciphertext $c$ we
    had $c=g_1\cdot g \cdot g_2+r$, where $r$ is the remainder, then for the same ciphertext we have $c=Ctip(g)
    g_1\cdot g'\cdot g_2+r$, where $r$ is again the remainder, and it coincides with the remainder of division of $c$
    by the initial $g$.
    \\
    \indent
    For even more confusion for decrypting system we may send "fake" ciphertext of the form $c''=c'+h$, where
    $c'$ is as above, and $h\in\mathbb{F}_q<X>$, such that $tip(g)$ does not divide any monomial in $h$. Note
    that such polynomials $h$ "incorporate" monomials from $NonTip(I)$, i.e. valid messages.
    A "plaintext" corresponding to $c''$ will be $p'+h$, which again gives rise to $g'$ as above. So, in our
    attack the "fake" ciphertext $c''$ contains either monomials divisible by $tip(g)$ and non-divisible. In
    addition, we note that the variety of such $c''$'s is very broad.
    \\
    \indent
    All considerations above imply that using private $G=\{g\}$ can be claimed as insecure. Note that right from the
    definition
    of a reduced Groebner basis we get that also private keys of the form $G=\{g_1,\dots,g_s\}$, where $G$ is the
    reduced Groebner basis for $I$ are also can be claimed as insecure, because we can apply out technique $s$
    times in order to be able to correctly decrypt valid ciphertexts. So, at this point only private keys of the form
    $G=\{g_1,\dots,g_s\}$, where $G$ is not reduced Groebner basis for $I$, and $s>1$, can give a hope on constructing
    a system not susceptible to chosen-ciphertext attacks.
    \\
    \indent
    As a final remark we would like to note that the same principle can be applied when cryptanalyzing the generalized
    (commutative) Polly Cracker cryptosystems (cf. section 2.1, [1]).

    \section{Conclusion}
    In this short note we have shown that newly proposed noncommutative Polly Cracker cryptosystem as it was worked
    through in [1] is susceptible to a chosen-ciphertext attack. This conceptually coincides with warnings stated in
    [4] as
    to using polynomial-based cryptosystems, and shows that more care should be put, when constructing such a system.

\end{document}